\def \Rmip {R_\textrm{mip}}
\def \rhoSM {\rho_\textrm{SM}}
\def \etaSM {\eta_\textrm{SM}}
\def \Emax {E_\text{max}}

\documentclass[11pt,twoside]{elsartmod}
\def\psnormal{\voffset=-0.7cm \headheight=0cm
\hoffset=0.cm \headsep=0.5cm 
\oddsidemargin=0.cm\evensidemargin=0cm \topmargin=0cm\parindent=1cm\footskip=1.5cm} 
\psnormal 
\usepackage{amssymb}
\usepackage{epsfig}
\usepackage{wrapfig}
\usepackage{amsmath}

\begin{document}

\begin{frontmatter}
% when double little pages, use these three lines
%\thispagestyle{empty}
%\mbox{}
%\pagebreak

\thispagestyle{empty}

%\begin{flushright}
%Draft \today
%\end{flushright}

\mbox{}
\vspace{1cm}
\mbox{}

\begin{center}
{\huge
\bf
Measurement of the $\mu$ decay spectrum
with the ICARUS liquid Argon TPC
}

\mbox{}
\vspace{0.5cm}
\mbox{}

{\Large
The ICARUS Collaboration
}

\newcommand{\Aname}[2]{#1}
\def\titlefoot#1{\centerline{\it #1}~\newline}
\renewcommand{\thefootnote}{\alph{footnote}}
\def\A{\kern+.6ex\lower.42ex\hbox{$\scriptstyle \iota$}\kern-1.20ex a}
\def\E{\kern+.5ex\lower.42ex\hbox{$\scriptstyle \iota$}\kern-1.10ex e}

\author[Napoli]{S.~Amoruso}{,}
\author[Aquila]{M.~Antonello}{,}
\author[LNGS]{P.~Aprili}{,}
\author[LNGS]{F.~Arneodo}{,}
\author[ETH]{A.~Badertscher}{,}
\author[Padova]{B.~Baiboussinov}{,}
\author[Padova]{M.~Baldo Ceolin}{,}
\author[Milano]{G.~Battistoni}{,}
\author[katowice]{B.~Bekman}{,}
\author[Pavia]{P.~Benetti}{,}
% ...
\address[Napoli]{\scriptsize Universit\`a Federico II di Napoli e INFN, Napoli, Italy}
\address[Aquila]{\scriptsize Universit\`a dell'Aquila e INFN, L'Aquila, Italy}
\address[LNGS] {\scriptsize INFN - Laboratori Nazionali del Gran Sasso, Assergi, Italy}
\address[ETH]{\scriptsize Institute for Particle Physics, ETH H\"onggerberg, Z\"urich, Switzerland}
\address[Padova]{\scriptsize Universit\`a di Padova e INFN,  Padova, Italy}
\address[Milano]{\scriptsize Universit\`a di Milano e INFN, Milano, Italy}
\address[katowice]{\scriptsize Institute of Physics, University of Silesia, Katowice, Poland}
\address[Pavia]{\scriptsize Universit\`a di Pavia e INFN, Pavia,  Italy}
% ...
\end{center}
\vspace{-0.6cm}

\begin{abstract}
Examples are given which prove the ICARUS detector quality through
relevant physics measurements. We study the $\mu$ decay energy
spectrum from a sample of stopping $\mu$ events acquired during the
test run of the ICARUS T600 detector. This detector allows the spatial
reconstruction of the events with fine granularity, hence, the precise
measurement of the range and $dE/dx$ of the $\mu$ with high sampling
rate. This information is used to compute the calibration factors
needed for the full calorimetric reconstruction of the events. The
Michel $\rho$ parameter is then measured by comparison of the
experimental and Monte Carlo simulated $\mu$ decay spectra, obtaining
$\rho = 0.72\pm 0.06
\textrm{ (stat.)} \pm 0.08 \textrm{ (syst.)}$. The energy resolution
for electrons below $\sim 50$~MeV is finally extracted from the
simulated sample, obtaining
$(E^e_\textrm{meas}-E^e_\text{MC})/E^e_{MC} = 11\%
/\sqrt{E\textrm{ [MeV]}} \oplus 2\%$.
\end{abstract}

\end{frontmatter}

\newpage
\setcounter{page}{1}

\section{Introduction}

The study of muon decay has played in the past a major role for the
understanding of weak interactions, being the only accessible purely
leptonic process. Muon decay was first described in a
model-independent way by Michel~\cite{MICHEL}, using the most general,
local, derivative-free, lepton-number conserving, four fermion
interaction. For unpolarized muons, the decay probability is given by:
\begin{equation}
\frac{dP}{dx}(x;\rho,\eta) = \frac{1}{N} x^2 \left(3 (1-x) +
\frac{2}{3}\rho(4x-3)+3\eta\frac{m_e}{\Emax} \frac{1-x}{x} +
\frac{1}{2} f(x) + \mathcal{O}(\frac{m_e^2}{\Emax^2})\right) 
\label{eq:michel}
\end{equation}
where $N$ is a normalization factor; $x=\frac{E_e}{\Emax}$ is the
\emph{reduced} energy (ranging from $m_e/\Emax$ to 1); $E_e$ and $m_e$
are respectively the total energy and mass of the electron produced in
the decay; $\Emax=m_\mu/2$ is the end-point of the spectrum; $f(x)$
is the term accounting for the first order radiative corrections
assuming a local V-A interaction~\cite{RADCORR}; finally, $\rho$ and
$\eta$ are the so-called Michel parameters, defined in terms of
bilinear combinations of the coupling constants of the general four
fermion interaction, and hence depending on the type of interaction
governing the decay process. For the Standard Model (SM) V-A
interaction, the parameters take the values $\rhoSM = 0.75$ and
$\etaSM = 0$.

\begin{figure}[!th]
\begin{center}
\epsfig{file=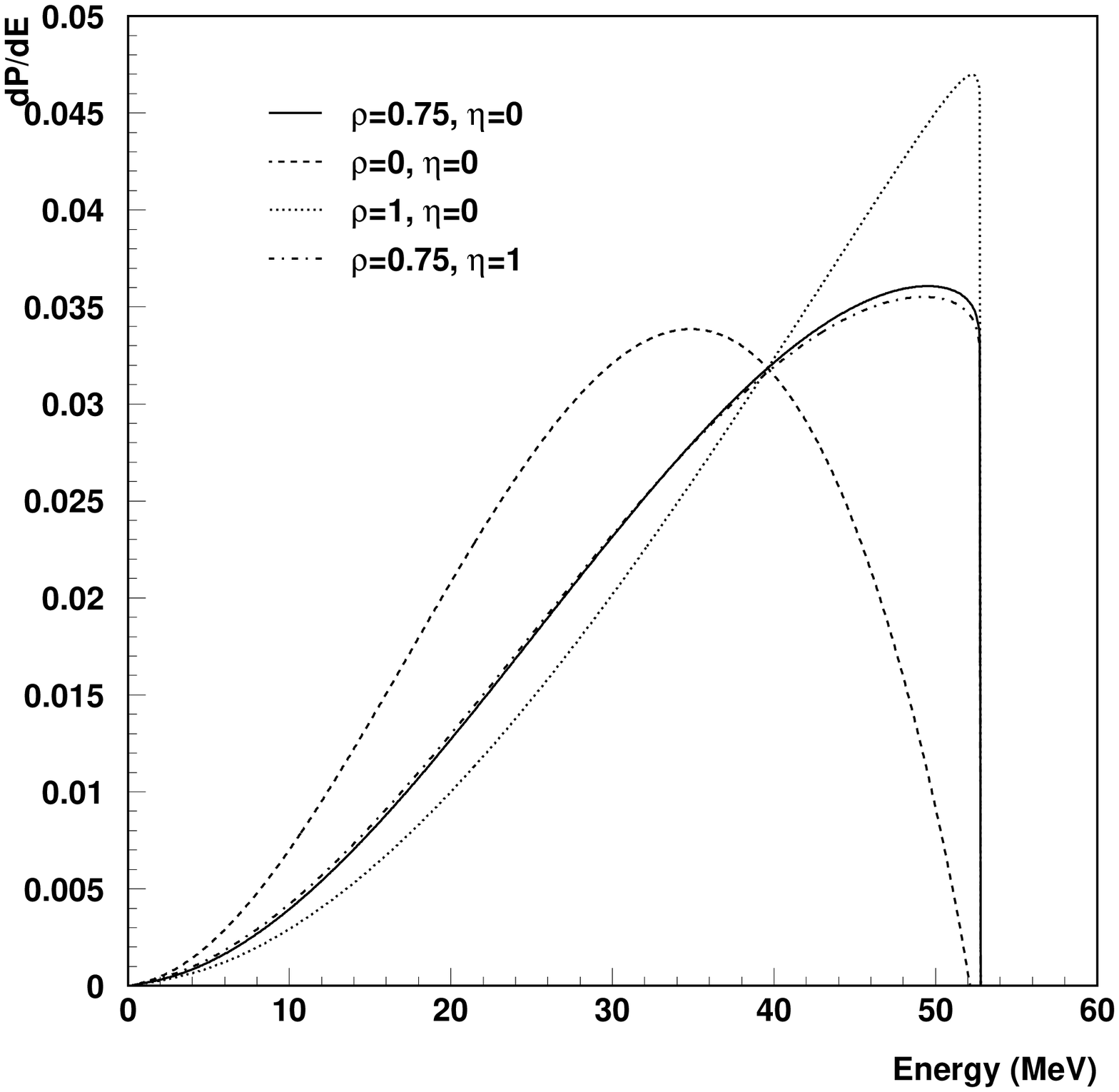,width=7.5cm}
\epsfig{file=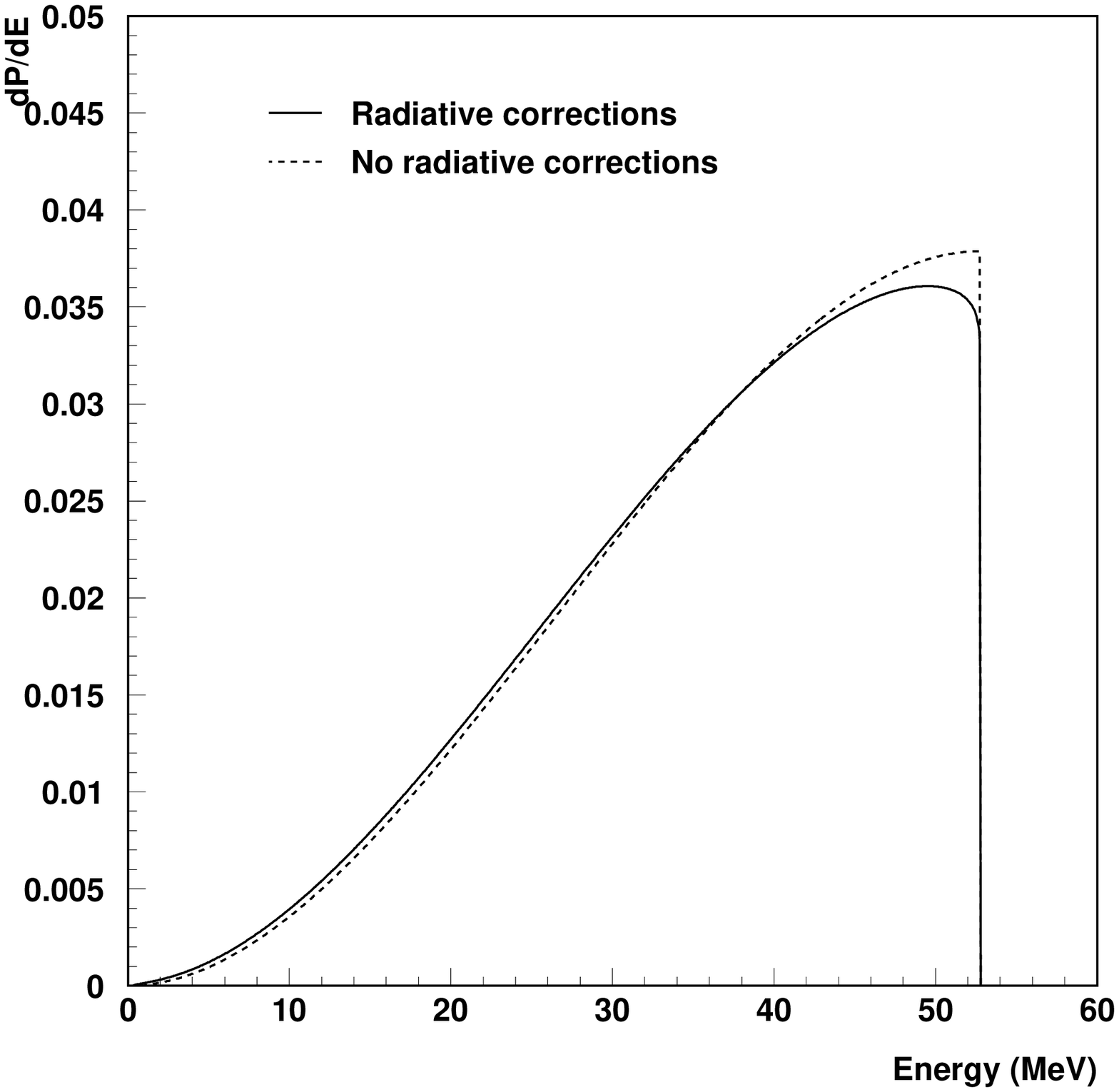,width=7.5cm}
\end{center}
\caption
{Left: Muon decay energy spectra for various values of the Michel
parameters including the V-A values. Right: Effect of the first order
radiative corrections on the muon decay spectrum ($\rho = 0.75$, $\eta
= 0$).}
\label{fig:puremichel}
\end{figure}

The V-A assumption has already been confirmed in muon decay with high
precision~\cite{FETSCHER,PDG} by the determination of the whole set of
Michel parameters and complementary
measurements. Figure~\ref{fig:puremichel} (left) shows the theoretical
shape of the $\mu$ decay spectrum for various values of the parameters
$\rho$ and $\eta$.  As shown in the figure, and expected from
inspection of Equation~\ref{eq:michel}, the shape of the spectrum is
more sensitive to $\rho$, since $\eta$ is weighted by $1/x$ and
hence determines the shape at low energy. The radiative corrections,
shown in Figure~\ref{fig:puremichel} (right), determine the shape of
the spectrum near the end-point and therefore the value of $\rho$ is
very sensitive to them. It has been shown that the overall effect of
the radiative corrections on the value of $\rho$ is of the order of
$6\%$~\cite{RADCORR}.

The Michel parameter $\rho$ has been measured in the past by several
groups (see Table~\ref{table:rhomeas}). Peoples~\cite{PEOPLES},
Sherwood~\cite{SHERWOOD} and Fryberger~\cite{FRYBERGER} have measured
$\rho$ in the late 60's using the high energy part of the $\mu$ decay
spectrum assuming the V-A value $\eta=0$. Derenzo~\cite{DERENZO} has
measured the lower part of the energy spectrum, and combined his data
with the previous results (essentially Peoples' measurement) into a
common two-parameter fit, to obtain $\rho$ with a precision of about
0.4$\%$, which is the most accurate existing measurement with no
assumptions for the value of $\eta$.  These results have been obtained
using dedicated experiments involving data samples of typically
several hundreds of thousands of events. More recently, the data from
electron-positron colliders have been used to measure the Michel
parameters of the purely leptonic $\tau$ decay near the $Z_0$
resonance~\cite{RHOTAU}. These measurements are based on the analysis
of samples that typically include several tens of thousands of
events. 

We present here a further measurement obtained with the ICARUS
detector, during its test phase (2001). ICARUS is a project, proposed
in 1985~\cite{ICARUS}, for the installation of a large liquid Argon
(LAr) time projection chamber (TPC) in the Gran Sasso Laboratory,
Italy, for the study of neutrino physics and matter
stability~\cite{CRUBBIA}. The physics potential of this type of
detector has been extensively described elsewhere, both for the final
project~\cite{ICARUS} and its initial phase with the $\sim 600$~t LAr
prototype (ICARUS T600)~\cite{T600PROP}.

The sample of events in which a muon enters, stops and eventually
decays in the detector's sensitive volume --hereafter called
\emph{stopping muon} sample-- constitutes an important benchmark to
evaluate the physics performance of ICARUS. Because of their simple
topology, stopping muon events are relatively easy to reconstruct in
space, allowing the computation of the different calibration factors
needed in the full calorimetric reconstruction. Thus, we can study the
muon decay spectrum and measure the Michel $\rho$ parameter, which
constitutes the first physics measurement performed with the novel
ICARUS detector technology, and proves that the technique is mature
enough to produce physics results. Our new result is not competitive
with those obtained from $\mu$ decay, and barely with those obtained
from $\tau$ decay. However, it must be remarked that this result has
been obtained using 1858 muon decay events with a non optimized
experiment. Our result stresses the capabilities of the ICARUS
technology to produce robust physics results.

\begin{table}[!t]
\begin{center}
\begin{tabular}{|c|r@{ $\pm$ }l|c|} \hline
Author & \multicolumn{2}{c|}{Value} & Assumption \\ \hline
Peoples       & 0.750 & 0.003 & $\eta\equiv 0$\\
Sherwood      & 0.760 & 0.009 & $\eta\equiv 0$\\ 
Fryberger     & 0.762 & 0.008 & $\eta\equiv 0$\\
Derenzo       & 0.752 & 0.003 & $-0.13 < \eta < 0.07$ \\
SLD	      & 0.72  & 0.09  $\pm$ 0.03 & lepton univers.\\
CLEO          & 0.747 & 0.010 $\pm$ 0.006 & lepton univers. \\
ARGUS         & 0.731 & 0.031 & lepton univers. \\
L3            & 0.72  & 0.04  $\pm$ 0.02 & lepton univers. \\
OPAL          & 0.78  & 0.03  $\pm$ 0.02 & lepton univers. \\
DELPHI        & 0.78  & 0.02  $\pm$ 0.02 & lepton univers. \\
ALEPH         & 0.742 & 0.016 & lepton univers. \\
This analysis & 0.72 & 0.06$\pm$ 0.08 & $-0.020 < \eta < 0.006$ \\
\hline
\end{tabular}
\caption{Results from previous measurements of the Michel $\rho$
parameter. First (second) quoted error is of statistical
(systematical) origin. Single error bounds correspond to statistical
and systematic errors added in quadrature.}
\label{table:rhomeas} 
\end{center}
\end{table} 

\section{Experimental setup}
\label{sec:setup}

ICARUS T600~\cite{T600} is a large cryostat divided in two identical,
adjacent half-modules of internal dimensions $3.6 \times 3.9 \times
19.9$ m$^3$, each containing more than 300~t of LAr. Each half-module
houses an internal detector composed of two TPC's (referred to as {\it
Left} and {\it Right} chambers), the field shaping system (race track
electrodes), monitors, probes, photo-multipliers, and is externally
surrounded by a set of thermal insulation layers. Each TPC is formed
by three parallel planes of wires, 3~mm apart, oriented at $0,
\pm60^\circ$ to the horizontal, with 3~mm pitch wires, positioned
along the left and right side walls of the half-module. The cathode
plane is parallel and equidistant to the wire planes of each TPC. A
high voltage system produces a uniform electric field (500~V/cm)
perpendicular to the wire planes, forcing the drift of the ionization
electrons (the maximal drift path is 1.5~m). The electric field is
uniform in the volume contained between the wire planes, the cathode
and the planes of race track electrodes (hereafter referred to as LAr
{\it active} volume).

The ionization electrons produced in the LAr active volume drift
perpendicularly to the wire planes due to the applied electric field,
inducing a signal ({\it hit}) on the wires near which they are
drifting while approaching the different wire planes. By appropriate
biasing, the first set of planes can be made non-destructive ({\it
Induction} planes), so that the charge is finally collected in the
last plane ({\it Collection} plane).  Each wire plane provides a
two-dimensional projection ({\it view}) of the event, where the
position in one coordinate is constrained by the hit wire, while the
signal timing with respect to the trigger{\rm } gives the position
along the drift direction. Each track is sampled by a large number of
wires. This is one of the main feature of the ICARUS technology,
exploited in the present analysis.

A full test of the T600 experimental set-up on the surface of the
earth was carried out in Pavia (Italy) during the period April-August
2001. One T600 half-module was fully instrumented to allow a complete
test under real experimental conditions. Although the test was mainly
intended as a technical run, a substantial amount of cosmic ray data
was acquired, since the detector was not shielded against cosmic rays
as will be the case in the Gran Sasso underground laboratory.  The
detector's conditions during data acquisition (DAQ) were not stable,
being subject to the optimization of various working parameters of the
detector.

One of the main goals of the test run was the detection of long muon
tracks crossing the detector at large zenith angles (horizontal
muons), extremely useful for an overall test of the detector
performance. Therefore, an external trigger system was set up to
select the horizontal muon events out of the overwhelming background
of atmospheric showers. The system consisted of two plastic
scintillator layers suitably positioned and arranged in a proper
coincidence trigger logic. In spite of the optimization of the trigger
for the acquisition of horizontal muon tracks, other kind of events
(in particular, stopping muon events) are expected to be acquired
within the DAQ time window. Most of these events can not be directly
correlated to the triggering event, hereafter they will be referred to as
\emph{out-of-time} events (conversely, those which are correlated to
the triggering event will be referred to as \emph{in-time} events).

\section{Data selection}

\begin{figure}[!tp]
\begin{center}
\epsfig{file=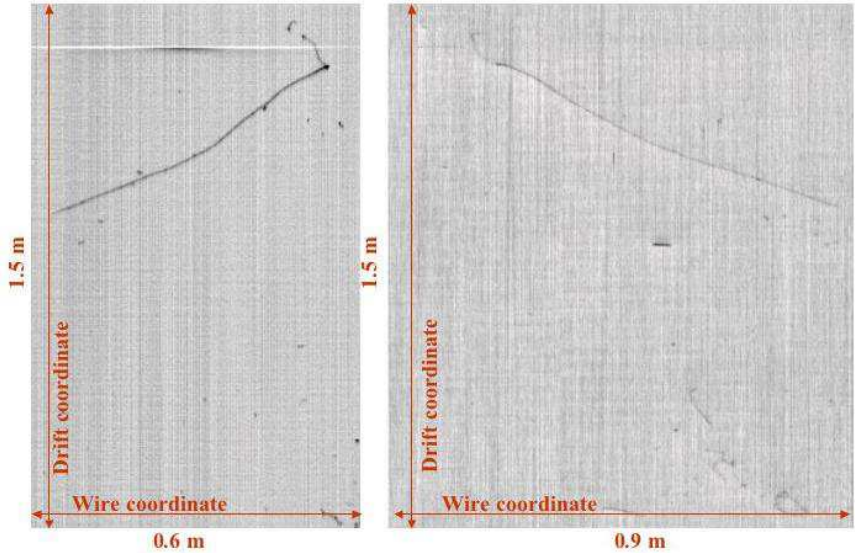,height=7.5cm}
\end{center}
\caption{Run 966 Event 8 Right chamber: muon decay event views
corresponding to the Collection (left) and second Induction (right)
wire planes.}
\label{fig:stopmu}
\end{figure}

Data selection was carried out by visual scanning using topological
criteria. Stopping muons can be recognized as minimum ionizing tracks
entering the detector, with increasing energy deposition and multiple
scattering angle when approaching the stopping point. Two different
topologies are expected depending on the process undergone by the
stopped muon, namely: decay or nuclear absorption. The probability of
the absorption processes has been evaluated to be 73$\%$ (0$\%$) for
negative (positive) muons~\cite{PRIMAKOFF}. For absorption events
($\mu^- + Ar \rightarrow Cl^*+\nu_\mu$), the excited Cl nucleus decays
emitting photons which then may interact via Compton scattering. For
decay events, a minimum ionizing electron track (length
shorter\footnote{The minimum observable track length is given by the
wire pitch: 3~mm.}  than $\sim$23~cm) follows the muon. In the present
analysis, both decay and absorption events have been used in the
determination of the calibration factors for the calorimetric
reconstruction (see section~\ref{sec:calrec}).
Figure~\ref{fig:stopmu} shows the two-dimensional projections
(produced by the Collection and the second Induction planes
respectively) of a typical muon decay event.

In the case of decay events, the muon and electron tracks are
separated by the muon stopping point, which is taken as the hit with
the maximal ionization. The stopping point is assigned to the muon
track. This may produce, for events where the muon and electron tracks
partially overlap, a loss of the first hits from the electron
track. This effect has been evaluated and introduced in the analysis
using a sample of simulated events, as explained in
section~\ref{sec:rho}.

A total of 5830 triggers were scanned, containing 4548 stopping muon
events. Out of them, 3370 (74$\%$) were further selected by the
preliminary quality cuts. Among the 1178 rejections, 45$\%$ of the
cases were events with deficiently measured drift charge since the
event happened in a region of the detector where the wires had an
inadequate polarization during the run; 25$\%$ of the cases were due
to an event occurring in a region with a substantial component of high
amplitude, correlated noise, which may fake the presence of an
ionizing track, hence, distorting the real event; the remaining 30$\%$
of rejections were due to several other effects, mainly events not
fully contained in the LAr sensitive volume\footnote{An event is
considered not to be fully contained if the muon or electron end
points (identified by the raise in the energy deposition per unit
length) are not present. No further fiducial cut on the LAr sensitive
volume is required.}, failures of the reconstruction program due to
complicated topologies or absence of one or more views due to problems
of the DAQ system.

\section{Data reconstruction}
\label{sec:recon}

\subsection{Spatial reconstruction}

A full 3D reconstruction of the selected muon events is first
performed. The spatial reconstruction of the muon tracks is needed in
order to compute the calibration factors entering the calorimetric
reconstruction of the events, as explained in
section~\ref{sec:calrec}.

A detailed description of the spatial reconstruction tools has been
reported elsewhere~\cite{T600,THESIS}. The 3D reconstruction is
performed in a hit basis using a three-step procedure:
\begin{enumerate}
\item
First, hits are searched for independently in every wire of each wire
plane. Hits are identified as signal regions of a certain width with
output values above the local mean. No information from adjacent wires
is used at this stage. A precise determination of the hit position and
charge are carried out by means of a fit using an analytical
function. 
\item
The identified hits are associated in the second step into
2-dimensional clusters of hits belonging to a common charge
deposition, such as tracks or showers. Clusters provide criteria for
the identification of the different patterns and for the
discrimination between signal and noise hits, based on the cluster hit
multiplicity.
\item
In the third step, the 3D coordinates of the hits from the different
reconstructed clusters are computed. Each wire plane constrains two
spatial degrees of freedom of the hits, one common to all the wire
planes (the drift coordinate) and one specific for each plane (the
wire coordinate). The redundancy on the drift coordinate allows the
association of hits from different views to a common energy
deposition, and together with the wire coordinates from at least two
planes, the determination of the hit spatial coordinates.
\end{enumerate}

\begin{figure}[!tp]
\begin{center}
\epsfig{file=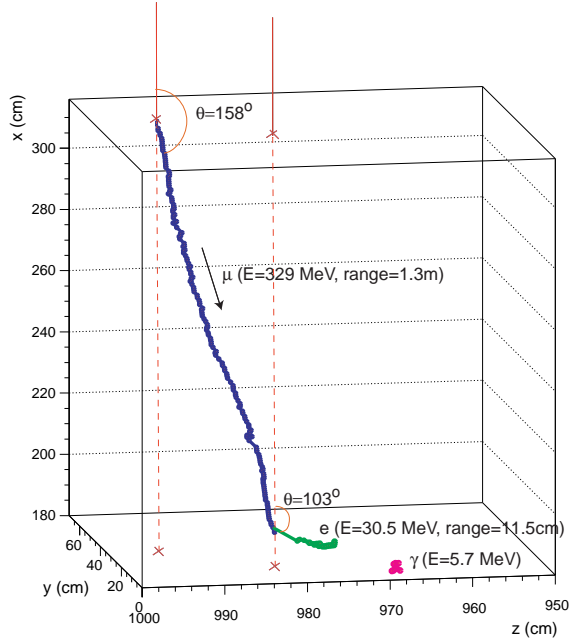,width=7.5cm}
\end{center}
\caption{Run 966 Event 8 Right chamber: fully reconstructed  muon
decay event.}
\label{fig:stopmu3d}
\end{figure}

Figure~\ref{fig:stopmu3d} shows a 3D reconstructed stopping muon
event. The muon decay topology is visible, with three reconstructed
tracks corresponding to the muon, the decay electron and a $\gamma-e$
conversion, respectively.

%For stopping muon events, the
%reconstruction program is seeded with the coordinates (in the
%wire/drift time plane) of the first hit of the muon track in at least
%two views, plus the coordinates of the muon and (eventually) the
%electron stopping points in one view. These coordinates define the
%different tracks (muon and electron) of the event. Moreover, the
%spatial reconstruction algorithm follows the identified hits along the
%muon track in at least two views, associating those corresponding to
%the same energy deposition by means of the common drift time
%coordinate. The muon track is finally
%\emph{smoothed} by approximating it by straight line segments, whose
%length is energy dependent and is optimized to get the best estimate
%of the muon energy loss and range.

\subsection{Calorimetric reconstruction}
\label{sec:calrec}

The ionization charge is precisely measured at the Collection wire
plane. The energy associated to a given hit is related to the
collected charge by means of
\begin{equation}
E = \frac{CW}{R}\  e^{(t-t_0)/\tau_e}\, Q
\label{eq:hitenergy}
\end{equation}
where $C=(152 \pm 2)\times 10^{-4}$~fC/(ADC$\times \mu$s) is
the calibration factor~\cite{CALIBRATION}; $W= 23.6^{+0.5}_{-0.3}
\textrm{ eV}$ is the average energy needed for the creation of
an electron-ion pair~\cite{MIYAJIMA}; $R$ is the electron-ion
recombination factor; ($t-t_0$) is the drift time of the electrons;
$\tau_e$ is the drift electron lifetime, which parametrizes the
attachment of drift electrons to impurities in LAr; and $Q$ is the
measured charge. $R$, $t_0$ and $\tau_e$ are extracted from the
reconstructed muon tracks, essentially by tuning them so that the
measured energy corresponds to the theoretical expectation for
stopping muons. This method determines the electron energy in a
bias-free way, since all the calibration parameters are tuned using
exclusively muon tracks. For the real experimental conditions $t_0$ is
given by the triggering system, but is unknown for most of the
stopping muon events used in this analysis, since they are out-of-time
events.

\begin{figure}[!tp]
\begin{center}
\epsfig{file=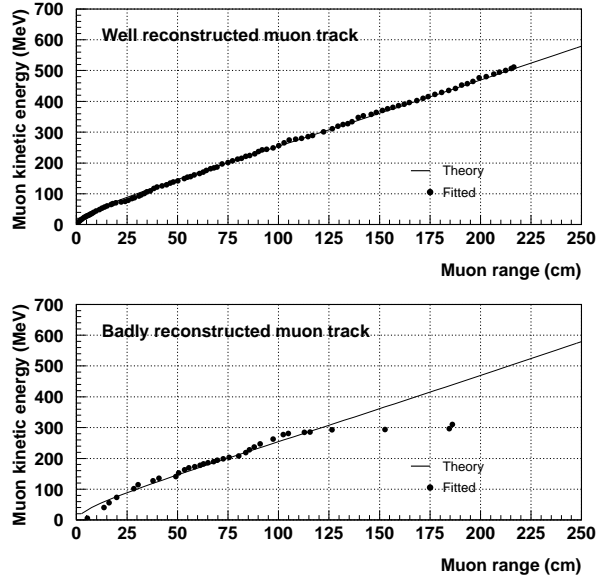,height=8.5cm}
\end{center}
\caption{Kinetic energy vs.\ range for typical well (top) and badly (bottom)
reconstructed muon tracks.}
\label{fig:good_bad}
\end{figure}

Only those tracks successfully reconstructed in space are used in the
determination of $\tau_e$ and $R$. A quantitative measurement of the
quality of the spatial reconstruction is provided by the goodness of
the fit of the muon track to the theoretical kinetic energy vs.\ range
curve, when $R$, $t_0$ and $\tau_e$ are left as free parameters.  A
total of 2690 stopping muon events are selected using this method. Two
examples of the energy vs.\ range curve measured for typical well and
badly reconstructed muon tracks are shown in
Figure~\ref{fig:good_bad}.

The drift electron lifetime, $\tau_e$, is measured for data taking
intervals of 24 hours. The measurement method is based on the
observation of the charge attenuation with the drift distance, and has
been published elsewhere~\cite{CALIBRATION}. However, two main
differences must be stressed~\cite{THESIS}. First, the charge of a hit
depends on the particle momentum, but also on the angle between the
track and the wire.
%energy of the different hits of a given stopping muon track depends on
%the track length and momentum associated to the hit. 
Therefore, in general the hit charges are not directly comparable. In
order to solve this problem, we normalize each charge with the
theoretically expected energy, computed with the Bethe-Bloch formula
given the range and length associated to the hit. Second, $t_0$ is
unknown for most of the events, and hence also the absolute drift
time/distance. In order to merge hits from different tracks with
different $t_0$ into a common measurement of $\tau_e$, we use the {\it
relative} charge attenuation (rather than {\it absolute}) as a
function of the drift time/distance. Using this method, we obtain
values of $\tau_e$ ranging from 1.20 to 1.70~ms (depending on the data
taking period), in agreement with the results previously
published~\cite{CALIBRATION}. Data are also compatible with a slight
dependence of the electron lifetime on the height: the electron
lifetime is about 15\% lower at the top than at the bottom of the
liquid argon volume, with an almost linear dependence on height.  The
measured charge is corrected for the attenuation using the value of
$\tau_e$ averaged for the LAr volume, therefore this decrease is
considered as an uncertainty when computing $R$ and introduces a
contribution to the total energy resolution $\left.\frac{\Delta
E}{E}\right|_{\tau_e} = 5\%$.

$R$ is computed as a function of the theoretical $dE/dx$ (estimated
from the range using the Bethe-Bloch formula), by comparing the
measured charge (corrected for the finite drift electron lifetime)
with the theoretical expectation for stopping muon
tracks~\cite{THESIS}. Using a sub-sample of 112 in-time events, for
which $t_0$ is known, we obtain $\Rmip = 0.640 \pm 0.013$ for minimum
ionizing particles, where the error is dominated by the uncertainties
of $\tau_e$ and of the length of the track segments used to evaluate
$dE/dx$. $\Rmip$ is used as a reference to tune the value of $t_0$ for
out-of-time events. The uncertainty of this procedure is estimated by
applying it to in-time events: the values of $t_0$ obtained for these
events are Gaussian distributed, with mean $2 \pm 9$~$\mu$s and width
$82 \pm 8$~$\mu$s. This can be interpreted as a negligible shift on
the energy scale ($< 0.3\%$) and an extra contribution to the energy
resolution of $\left.\frac{\Delta E}{E}\right|_{t_0} = 7\%$. However,
it must be stressed that this contribution is due exclusively to the
fact that the analyzed events are out-of-time, therefore it will not
be present for the detector's real experimental conditions. The whole
muon data sample can be used next to compute $R$ for higher values of
$dE/dx$ up to $\sim 5$~MeV/cm. A linear dependence of $R^{-1}$ on
$dE/dx$ with slope $0.11 \pm 0.01$~cm/MeV is found.

\section{Results}

\subsection{Determination of the Michel $\rho$ parameter}
\label{sec:rho}

\begin{figure}[!t] 
\begin{center}
\epsfig{file=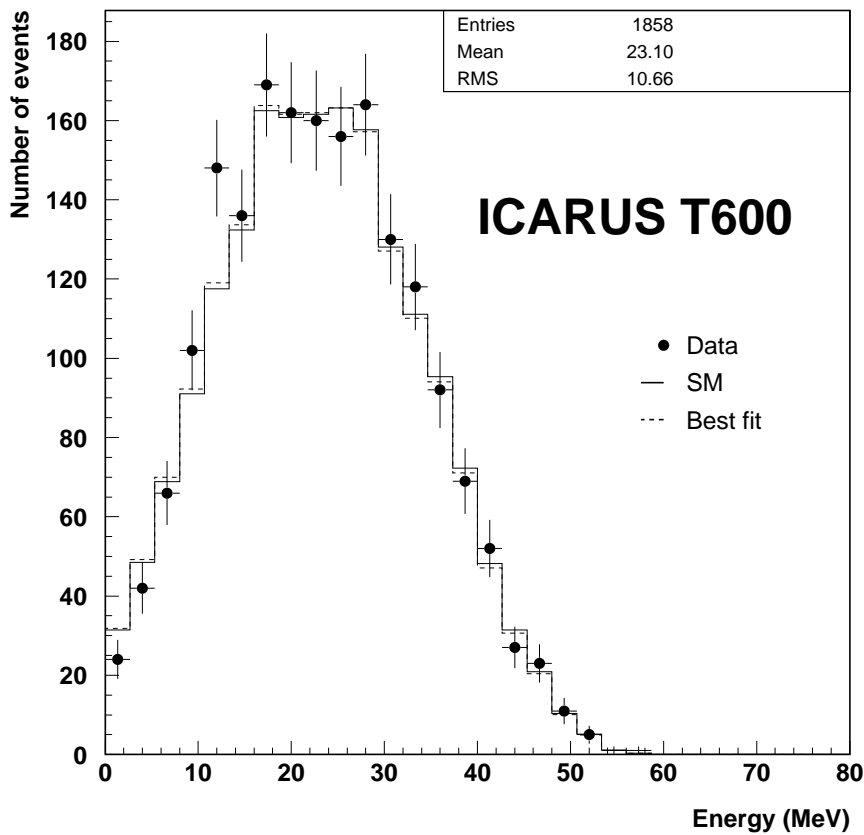,width=12cm}
\end{center} 
\caption{Energy spectrum of electrons from muon decays in the ICARUS T600 detector (electron
energy lost by ionization in LAr). The plot shows the measured
distribution (dots) compared both to the SM expectation (solid line)
and the best fit with $\rho$ and $\eta$ as free parameters (dashed
line).}
\label{fig:eletrackfit} 
\end{figure}

From the selected 2690 stopping muon events, a total of 1858 contain
electrons available for the determination of the Michel $\rho$
parameter (while the remaining ones are mainly absorption events).
The measured energy does not correspond directly to the $\mu$ decay
spectrum (Equation~\ref{eq:michel}) since no attempt to recover the
energy loss by Bremsstrahlung radiation has been carried out\footnote{
This does not decrease the statistical accuracy of the measurement and
avoids the problem of associating photons to the electron
track. Indeed, background conditions in data taking at the surface are
such that it was preferred not to worry about a selection of photons
pertaining to the electron.}. The spectrum corresponds to the fraction
of the electron energy lost by ionization in LAr. Therefore, in order
to measure the Michel $\rho$ parameter we have chosen an approach
based on the comparison with a Monte Carlo (MC) simulated event
sample. This sample has been generated using FLUKA~\cite{FLUKA} and
consists of $10\,000$ electrons from muon decay events inside the
detector's sensitive volume. The simulation includes all detector
effects except for the presence of impurities in LAr. MC events are
reconstructed using the same tools as for the data, and the effect of
the impurities and the determination of $t_0$ are included in average
by smearing the measured energy ($E$) using a Gaussian function of
width $\sigma=0.09 E$. The effect of the muon track (which is absent
in the MC sample) is the loss of some of the first hits of the
electron track due to the overlapping with the muon track.  The number
of hits of MC electron tracks is found to be on average 2.14 higher
than for data electron tracks. Thus, the effect of the muon track can
be included on average in the simulation by removing, when computing
the energy, the first 2 hits (3 in 14$\%$ of the cases) of the
electron track. The measured and simulated energy spectra are compared
in Figure~\ref{fig:eletrackfit}, and they are found to be in good
agreement ($\chi^2/ndf = 14.0/20$).

\begin{figure}[!t] 
\begin{center}
\epsfig{file=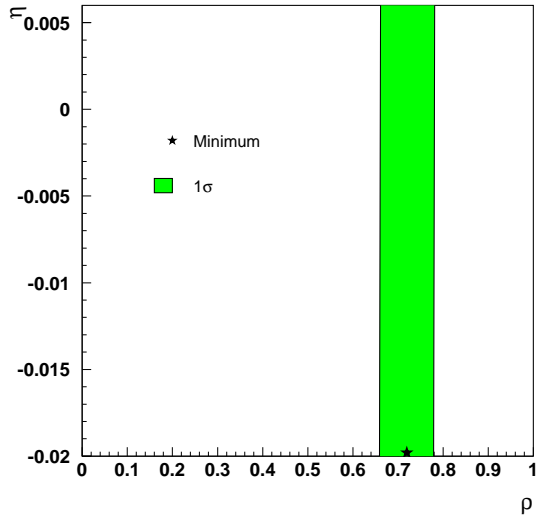,width=8cm}
\end{center} 
\caption{Results from the fit of the Michel electron spectra. The dot
and band show, respectively, the position of the minimum and 1$\sigma$
region within the considered parameter's space.}
\label{fig:scatter} 
\end{figure}

Given the greater sensitivity of the shape of the spectrum to $\rho$
than to $\eta$, and the level of correlation between these two
parameters (see Equation~\ref{eq:michel}), we measure $\rho$ while
constraining $\eta$ within its experimentally allowed interval
($-0.020 < \eta < 0.006$)~\cite{ETA}. Since the MC sample is generated
using the SM values ($\rhoSM$ and $\etaSM$), the spectrum for an
arbitrary pair of values, $\rho$ and $\eta$, is built by weighting
each MC event with a factor:
\begin{equation}
w = \frac{\frac{dP}{dx}(x_\text{MC};\rho,\eta)}
	{\frac{dP}{dx}(x_\text{MC};\rhoSM,\etaSM)}
\end{equation}
where $x_\textrm{MC}=E_\textrm{MC}/E_\textrm{max}$, and
$E_\textrm{MC}$ is the generated energy. We extract the value of
$\rho$ as that for which the best fit between the simulated and
measured energy spectra is obtained, which yields $\rho = 0.72 \pm
0.06$, where the error has a statistical origin and includes the
correlation with $\eta$. Figure~\ref{fig:scatter} shows the best fit
point and the 1$\sigma$ region of the fit within the considered Michel
$\rho$-$\eta$ plane. As expected, our measurement has no sensitivity
for the determination of $\eta$ within the considered experimental
bounds, and yields a stable value of $\rho$ for the whole allowed
$\eta$ range.

There are two types of systematic uncertainties that can affect this
measurement, namely: a bad estimate of the energy resolution and a
systematic shift in the global energy scale. Using the MC sample, the
uncertainty of $\rho$ is estimated to be 0.01 (0.03) for an extra $\pm
5\%$ ($\pm 10\%$) contribution to the energy resolution, and 0.04
(0.08) for a shift in the energy scale of $\pm 1\%$ ($\pm 2\%$). The
energy resolution is determined essentially by the wire's
signal-to-noise (S/N) ratio and the additional contributions from the
drift electron lifetime, recombination and $t_0$ corrections. All
these contributions have been extracted from the data and included in
the MC sample, and its uncertainty is estimated to be smaller than
5$\%$. Furthermore, the effect on the final result becomes negligible
by removing the last two bins of the electron spectrum when performing
the fit. Conversely, the absolute energy scale is given by $R$, which
is determined for minimum ionizing particles with an accuracy of 2$\%$
including systematic effects (see
section~\ref{sec:calrec}). Therefore, the total error is dominated at
this level by the systematics due to the uncertainty in the energy
scale. With a more precise calibration method, which will be available
in the real experimental conditions, it is reasonable to think that
the absolute energy scale will be known to the $1\%$ level, hence
reducing the systematic error of $\rho$ to 0.04. Conversely, with the
calibration method used in this analysis, the error of the absolute
energy scale is dominated by the systematics on the determination of
the length of the muon track segments. Therefore, no further
improvement of the result can be foreseen by using a larger data
sample.  The final result is:
\begin{equation}
\rho = 0.72\pm 0.06 \textrm{ (stat.)} \pm 0.08 \textrm{ (syst.)}
\end{equation}
which is compatible with the V-A value.

\subsection{Energy resolution}

The high level of agreement between the experimental and the generated
muon decay spectra allows us to estimate the detector's energy
resolution using the simulated sample. We have evaluated separately
the contributions to the final energy resolution coming from the
following sources:
\begin{itemize}
\item 
Electronic noise: This contribution depends on the S/N ratio. For
non-correlated noise and constant S/N ratio (which is a good
approximation of the real experimental conditions) this contribution
is expected to vanish at high energies (i.e.\ for high numbers of
wires).

\item
Reconstruction effects: This contribution is intrinsic to the signal
extraction method we use, and independent of the energy. The main
contribution comes from hits whose shape is not well reproduced by the
fitting function. Such a case may arise for tracks with small angle
with respect to the drift direction, or unresolved close
hits~\cite{THESIS}. Other minor contributions are due to undetected
hits or fake hits created during the automatic reconstruction
procedure.

\item
Calorimetric calibration: As already mentioned in
Section~\ref{sec:calrec}, an extra contribution to the energy
resolution is expected when correcting for the electron attachment to
impurities using an average value of the drift electron lifetime. This
contribution is also energy independent and has been estimated in
Section~\ref{sec:calrec}. In the case of this particular analysis, we
have also considered an extra contribution to the energy resolution
coming from the determination of $t_0$ for the different events.
\end{itemize}

\begin{figure}[!t] 
\begin{center}
\epsfig{file=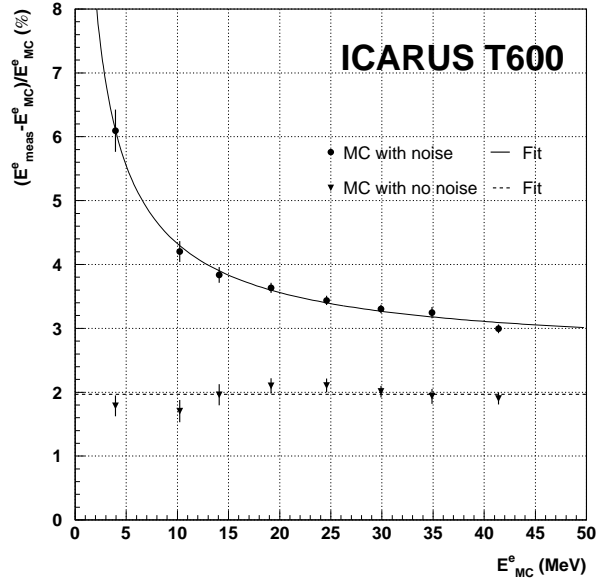,width=8.5cm}
\end{center} 
\caption{
Energy resolution as a function of the electron energy.}
\label{fig:trgauss}
\end{figure}

The first two contributions have been evaluated with MC samples
generated in the appropriate conditions (see
Figure~\ref{fig:trgauss}). In both cases, the energy resolution is
defined as $(E^e_\textrm{meas}-E^e_\text{MC})/E^e_\text{MC}$, where
$E^e_\textrm{MC}$ and $E^e_\text{meas}$ are, respectively, the MC
generated and measured energy of the electron track, disregarding
bremsstrahlung losses (i.e.\ only the fraction of the energy
lost by the electron by ionization of Ar atoms is compared). For every
$E^e_\text{MC}$ bin, the central value and error of the resolution are
obtained from a Gaussian fit, with errors of a statistical nature.

In order to evaluate the energy resolution expected from the
reconstruction effects, we have generated the MC sample with no
electronic noise, and passed it through the whole reconstruction
chain. In such a case, we obtain a constant resolution of $(1.97 \pm
0.05)\%$ (see Figure~\ref{fig:trgauss}). Conversely, the joint
effect of the electronic noise and the reconstruction is evaluated
with the MC sample generated including the electronic noise. A fit to
the points is performed by the function
$(E^e_\textrm{meas}-E^e_\text{MC})/E^e_\text{MC} = a/\sqrt{E\textrm{
[MeV]}} \oplus b$, where the best fit is obtained for (errors are of
statistical origin) $a = (11 \pm 1)\%$ and $b = (2.5 \pm 0.3)\%$ (see
Figure~\ref{fig:trgauss}). This result is compatible with the
assumption of a vanishing contribution of the electronic noise at high
energies.

\section{Conclusions}

We have performed the first physics measurement with the ICARUS LAr
TPC detection technique: the determination of the Michel $\rho$
parameter from the detailed study of the muon decay spectrum using the
stopping muon event sample from the ICARUS T600 detector test run with
cosmic rays. We obtain $\rho = 0.72\pm 0.06 \textrm{ (stat.)} \pm 0.08
\textrm{ (syst.)}$, in agreement with the SM value $\rho = 0.75$. This
measurement involves the exploitation of both spatial and calorimetric
reconstruction capabilities of the detector. Therefore, the obtained
result constitutes a proof of the maturity of the detection
technique to produce high quality physics results, in view of the
operation of the detector in the Gran Sasso underground laboratory. We
have also estimated the energy resolution including detector and
reconstruction effects from a sample of MC simulated Michel
electrons, obtaining $(E^e_\textrm{meas}-E^e_\text{MC})/E^e_{MC}$
compatible with $11\%/\sqrt{E\textrm{ [MeV]}} \oplus 2\%$, where the
raise at low energies is due to the electronic noise, and the constant
term arises from reconstruction effects.

\section*{Acknowledgments}

We warmly acknowledge N. Makrouchina, F. Varanini and S. Levorato for
their contribution to the scanning of the stopping muon events.  In
addition, we would like to warmly thank the many technical
collaborators that contributed to the construction of the T600
detector and to its operation.  We are glad of the financial and
technical support of our funding agencies and in particular of the
Istituto Nazionale di Fisica Nucleare (INFN), of ETH Z\"urich and of
the Fonds National Suisse de la Recherche Scientifique, Switzerland.
The Polish groups acknowledge the support of the State Committee for
Scientific Research in Poland, 2P03B09520, 2P03B13622,
105,160,620,621/E-344,E-340,E-77,E-78/SPS/ICARUS/P-03/DZ211-214/2003-2005;
the INFN, FAI program; the EU Commission, TARI-HPRI-CT-2001-00149.
The Spanish group is supported by the Ministry of Science and
Technology (project FPA2002-01835).

\end{document}